\documentclass[prl,showpacs,superscriptaddress,aps,floatfix,twocolumn]{revtex4}

\usepackage{color}
\usepackage{graphicx}
\usepackage{bm}
\usepackage{amsmath}
\usepackage{amsfonts}
\usepackage[centerlast]{subfigure}

\newcommand{\St}{\mathcal{S}}

\begin{document}

\title{Clustering of heavy particles in the inertial range of turbulence}

\author{J.\ Bec} \affiliation{CNRS UMR\,6202, Observatoire de la C\^ote
  d'Azur, BP4229, 06304 Nice Cedex 4, France.}

\author{M.\ Cencini} \affiliation{CNR, Istituto dei Sistemi Complessi,
  Via dei Taurini 19, 00185 Roma, Italy.} \affiliation{INFM-SMC c/o
  Dip.\ di Fisica Universit\`a Roma 1, P.zzle A.\ Moro 2, 00185 Roma,
  Italy.}

\author{R.\ Hillerbrand} \affiliation{CNRS UMR\,6202, Observatoire de la
  C\^ote d'Azur, BP4229, 06304 Nice Cedex 4, France.}
  \affiliation{Institut f\"ur theoretische Physik, Westf\"alische
  Wilhelms-Univ., M\"unster, Germany.}

\begin{abstract}
  A statistical description of heavy particles suspended in
  incompressible rough self-similar flows is developed.  It is shown
  that, differently from smooth flows, particles do not form fractal
  clusters. They rather distribute inhomogeneously with a statistics
  that only depends on a local Stokes number, given by the ratio
  between the particles' response time and the turnover time
  associated to the observation scale. Particle clustering is reduced
  when increasing the fluid roughness. Heuristic arguments supported
  by numerics are used to explain this effect in terms of the
  algebraic tails of the probability density function of the velocity
  difference between two particles. These tails are a signature of
  events during which particle couples approach each other very
  closely.
\end{abstract}

\date{\today}


\maketitle

In turbulent flow, finite-size particles heavier than the carrier
fluid, as e.\,g.\ water droplets in air, possess inertia and have a
finite response time to the fluid motion.  Thus their dynamics
markedly differs from that of simple tracers and, in particular, such
inertial particles distribute in a strongly inhomogeneous manner even
if the underlying flow is incompressible.  Modelling these
fluctuations in the particle concentration is important in
engineering~\cite{cst98}, cloud physics~\cite{pruppacher}, and
planetology~\cite{depater}.

The turbulent motion of the carrier fluid spans many active spatial
and temporal scales~\cite{frisch}. Below the Kolmogorov length-scale,
viscous dissipation dominates; the velocity field is differentiable
and is characterized by a single time scale. There the motion of
inertial particles is governed by the fluid strain and their
dissipative dynamics leads their trajectories to converge to a
dynamically evolving attractor.  For any given response time of the
particles, their mass distribution is singular and generically
scale-invariant with multifractal properties at small
scales~\cite{ekr96,bff01,b05}. Above the Kolmogorov scale, the fluid
velocity field is not smooth anymore but, according to the Kolmogorov
1941 theory, almost self-similar with H\"older exponent $h =
1/3$~\cite{frisch}.  This so-called inertial range is characterized by
a broad-spectrum of time scales. However, the finite response time of
the suspended particles introduces a new scale, breaking
self-similarity in the particle distribution. This is consistent with
the observation that particles typically concentrate on different
scales with the largest deviation from uniformity arising when their
response time is of the order of the eddy turnover
time~\cite{ef94,ffs03}. As already noticed in \cite{bff01}, these
deviations are not scale-invariant and have a different origin from
those observed in the viscous range of turbulence.  With few
exceptions~\cite{ss02,bdg04,fss05}, clustering in the inertial range
received considerably less attention than small-scale clustering.

In this Letter we focus on second-order statistics of the particle
distribution at scales within the inertial range. These statistics can
be completely described in terms of the relative motion of particle
pairs.  Within the inertial range, two concurrent mechanisms
responsible for particle clustering can be identified: a dissipative
dynamics due to their viscous drag with the fluid and ejection from
persistent vortical regions by centrifugal forces~\cite{maxey}. By
modelling the carrier flow as a rough, self-similar random velocity
field which is $\delta$-correlated in time, we eliminate the latter
effect: the absence of any persistent structure in the considered
carrier flow ensures that centrifugal forces play no role. This model
pertains to {\em very} heavy particles whose response time is much
larger than the typical correlation time of the ambient
fluid~\cite{bch06,fh06}. We show heuristically and numerically that
the scale-invariance of the velocity field does not extend to the
particle distribution, and that roughness of the carrier velocity
weakens clustering.  This effect is explained by the dependence of the
relative velocity distribution on the fluid velocity H\"older
exponent.

\begin{figure*}[ht]
  \centerline{ \includegraphics[height=0.35\textwidth]{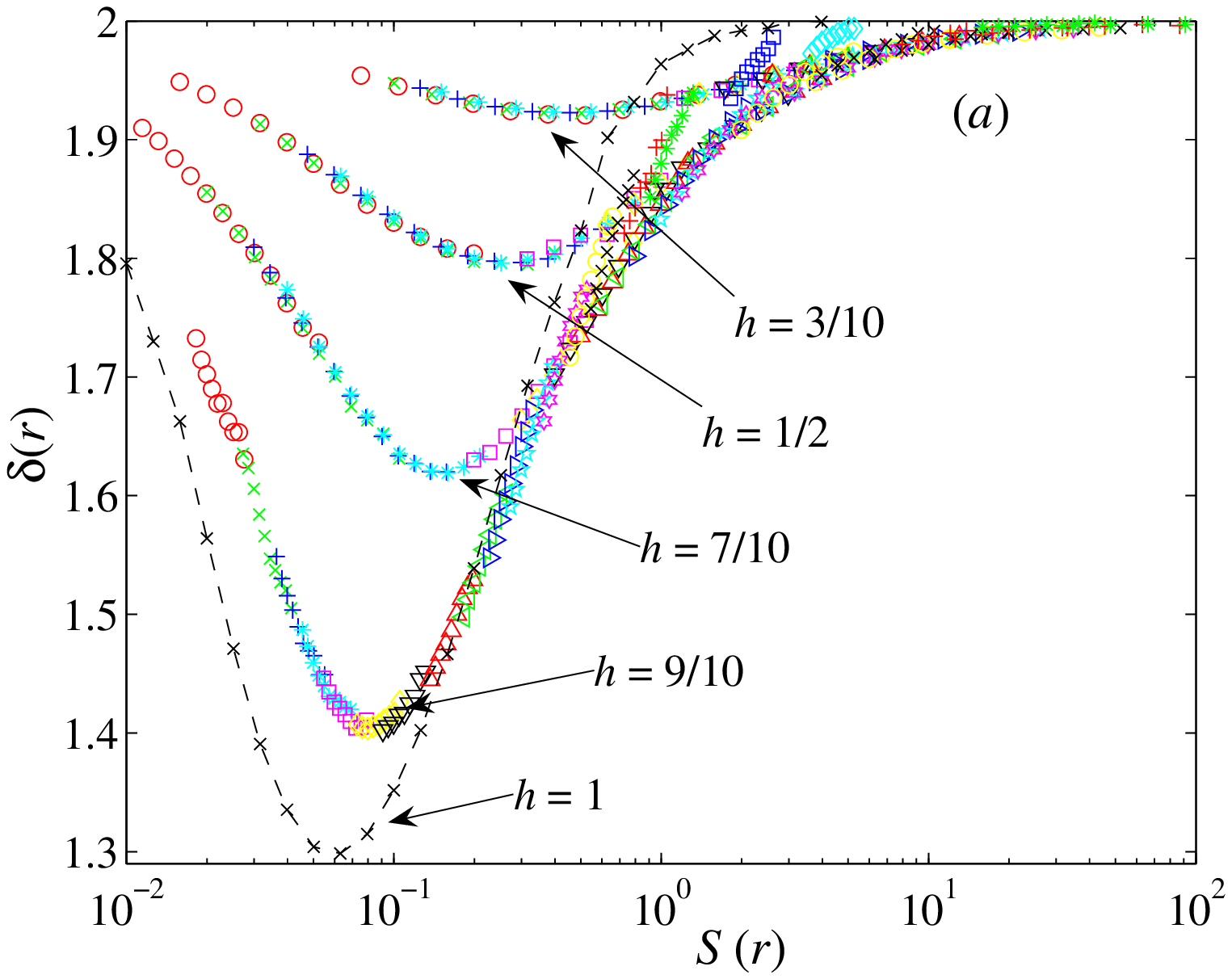}
  \qquad \includegraphics[height=0.35\textwidth]{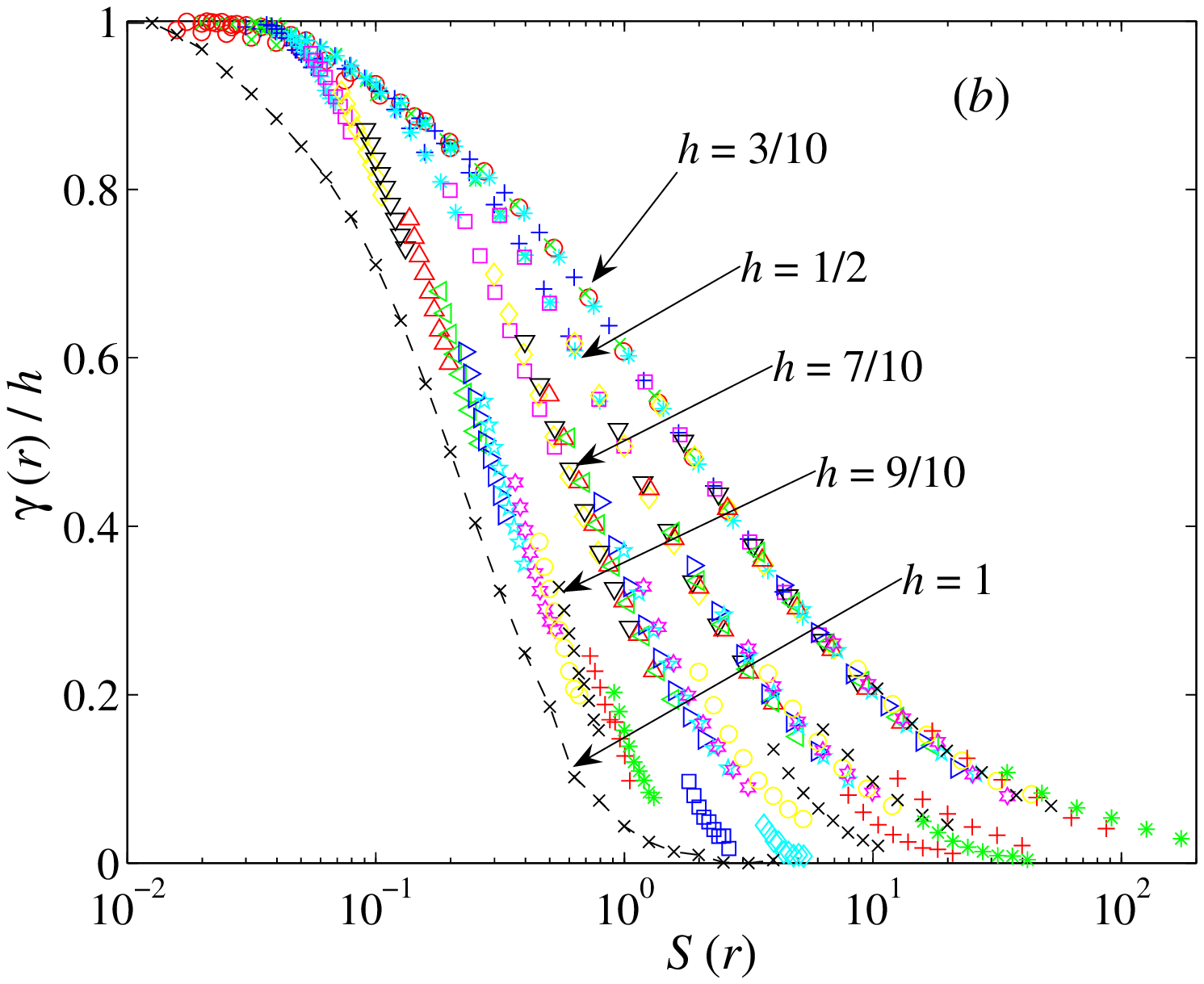}}
  \vspace{-10pt}
  \caption{(\textit{a}) Local correlation dimension $\delta(r)$ versus
    the scale-dependent Stokes number $\St(r)=D_1\tau/r^{2(1-h)}$ for
    various values of the H\"older exponent $h$ in two dimensions
    $d=2$.  (\textit{b}) Ratio between the local exponent $\gamma(r)$
    of the particle velocity and $h$ versus $\St(r)$. Different
    symbols (colors online) refer to different values of the particle
    response time $\tau$.\vspace{-10pt}}
  \label{fig:pdfZ}
\end{figure*}

The relative motion of two particles is described by their separation
  $\bm R$ that obeys the equation \cite{bch06,maxey}
\begin{equation}
  \ddot{\bm R} = -\frac{1}{\tau}\left[\dot{\bm R} - \delta\bm u (\bm
  R, t) \right]\,,
  \label{eqn:eomRV}
\end{equation}
where the dots denote time derivatives, $\tau$ is the Stokes time, and
$\delta\bm u(\bm r,t)=\bm u(\bm x+\bm r,t)- \bm u(\bm x,t)$ is the
fluid velocity difference.  The velocity $\bm u$ is assumed to be a
stationary, homogeneous and isotropic Gaussian field with correlation
\begin{equation}
  \langle u_i(\bm x, t) \, u_j(\bm x^\prime, t^\prime) \rangle =
 2D_0\delta_{ij} - B_{ij} (\bm x-\bm x^\prime)\,\delta(t- t^\prime)
 \,,
  \label{eq:correldeltau}
\end{equation}
where $D_0$ is the velocity variance.
For rough self-similar flows, the function $B$ takes the form
\begin{equation}
B_{ij} (\bm r) = D_1\, r^{2h} \left[ (d-1+2h)\,\delta_{ij} -
  2h\,{r_ir_j}/{r^2} \right]\,,
  \label{eq:corr}
\end{equation}
where $r=|\bm r|$, $d$ is the space dimension, $h \in [0,1]$ the
H\"older exponent of the carrier velocity field, and $D_1$ a constant
measuring the intensity of turbulence. This kind of velocity field was
introduced by Kraichnan~\cite{k68} to model passive scalar transport.

By defining $s = t/ \tau$ and rescaling $\bm R$ by the observation
scale $\ell$, it is easily seen that the above dynamics, and thus all
the statistical properties of particle pairs at scale $\ell$ only
depend on the {\em local Stokes number} $\St(\ell) = D_1\tau /
\ell^{2(1-h)}$. This dimensionless quantity, first introduced
in~\cite{ffs03}, is the ratio between the particle response time
$\tau$ and the turnover time at scale $\ell$. It measures the
scale-dependent effects of inertia.  At large scales ($\ell\to\infty$)
inertia becomes negligible ($\St(\ell)\to0$) and particles recover the
incompressible dynamics of tracers. Conversely, since
$\St(\ell)\to\infty$ for $\ell\to 0$, inertia effects become dominant
at small scales and the dynamics approaches that of free particles.
For both very large and very small values of the Stokes number, the
particles distribute uniformly in space. Strong inhomogeneities appear
for $\St(\ell)\approx 1$.

Note that in unbounded carrier flows, the separation between two
particles asymptotically grow indefinitely with time and thus the
dynamics (\ref{eqn:eomRV}) never reaches a statistical steady
state. However, real turbulent flows are bounded, allowing for
statistical equilibrium to be reached. Boundary conditions are thus
implemented in the considered model by imposing, for instance,
reflection of the inter-particle distance at $|\bm R| = L$.  Clearly,
since self-similarity is broken by the presence of boundaries, the
aforementioned scaling arguments apply only at scales $\ell\ll L$.

For smooth carrier flows ($h=1$), there is a unique time scale so that
the dynamics only depends on the global Stokes number
$\St=D_1\tau$. Inhomogeneities in the particle distribution can be
quantified by $d-\mathcal{D}_2$, where $\mathcal{D}_2$ is the
\emph{correlation dimension}
\begin{equation}
  \mathcal{D}_2 = \lim_{r\to0} \delta(r),\quad
  \delta(r) = {\rm d} \left(\ln P_2(r)\right)/{\rm d} \left(\ln r\right)\,,
  \label{eq:DefCorrDim}
\end{equation}
$P_2(r)$ denoting the probability that $|\bm R| < r$. In
$\delta$-correlated smooth flows, just as in real suspensions, the
correlation dimension $\mathcal{D}_2$ non-trivially depends on the
Stokes number~\cite{bch06}.

For non-smooth but H\"older-continuous flows, $\mathcal{D}_2=d$ for
{\em all} particle response times. Information on the inhomogeneities
of the particle distribution is entailed in the local correlation
dimension $\delta(r)$ defined in (\ref{eq:DefCorrDim}). From above
arguments, $\delta(r)$ is expected to depend only on $h$ and on the
local Stokes number $\St(r)$ when $r\ll L$.  This is confirmed
numerically for $d=2$ as shown in Fig.~\ref{fig:pdfZ}a, where the
local dimension $\delta(r)$ is represented as a function of $\St(r)$.
The collapse of the data for various response times $\tau$
demonstrates the dependence on the local Stokes number.  Comparing
various values of the exponent $h$, we observe that when the fluid
becomes rougher, the intensity of clustering weakens. In particular,
the minimum of $\delta(r)$ gets closer to $d=2$ as $h$ decreases.
Notice that for $h=1$, the Stokes number $\St$ does not depend on $r$
and data refer to the actual value of the correlation dimension, which
is well defined for smooth flows (see \cite{bch06} for details).

We now turn to the typical velocity difference $\dot{\bm R}$ between
two particles and its dependency on the separation $\bm R$.  For
smooth flows, when $|\bm R|\to 0$ an algebraic behavior of the form
$|\dot{\bm R}|\sim |\bm R|^\gamma$ is observed, defining a H\"older
exponent $\gamma$ for the particle velocities.  This exponent
decreases from $\gamma = h = 1$ for $\St=0$, corresponding to a
differentiable particle velocity field, to $\gamma = 0$ for
$\St\to\infty$, that means particle moving with uncorrelated
velocities ~\cite{bch06}.  In non-smooth flows the exponent $\gamma$
is asymptotically equal to the fluid H\"older exponent $h$ at large
scales ($\St(r)\to 0$), while it approaches $0$ at very small scales
($\St(r)\to\infty$). Therefore, similarly to $\delta(r)$, all relevant
information is entailed in the scale dependence of the local exponent
$\gamma(r)$.  As for the local correlation dimension, this exponent
only depends on the fluid H\"older exponent and on the local Stokes
number; this is confirmed in Fig.~\ref{fig:pdfZ}b, showing the ratio
$\gamma(r)/h$ versus $\St(r)$ for various values of $h$. It is worth
noticing that the transition from $\gamma(r) = h$ to $\gamma(r)=0$
shifts towards larger values of the local Stokes number and broadens
as $h$ decreases.  The fact that $\gamma(r)=h$ for $r\to \infty$
implies that the particles should asymptotically experience Richardson
diffusion~\cite{frisch} just as simple tracers.

To get a deeper understanding of the mechanisms of clustering, we
transform the equations of motion
(\ref{eqn:eomRV})-(\ref{eq:correldeltau}) into a system of stochastic
differential equations with additive noise.  Adapting to rough flows
the strategy first proposed in \cite{p02} for smooth carrier
velocities fields and used in \cite{mw04}, we make the following
change of variables:
\begin{eqnarray}
  X &=& (\tau/L^2) \, (|\bm R|/L)^{-(1+h)}\, \bm R\cdot \dot{\bm R}
  \,,
  \nonumber
  \\
  Y &=& (\tau/L^2) \, (|\bm R|/L)^{-(1+h)}\, |\bm R
  \wedge\dot{\bm R} |
  \,,
  \nonumber
  \\
  Z &=& \displaystyle (|\bm
  R|/L)^{1-h}\,. \label{eq:changevar}
\end{eqnarray}
The variables $X$ and $Y$ refer to the longitudinal and the transverse
dimensionless velocity differences, respectively.  When $d=2$, the
dynamics (\ref{eqn:eomRV}) in the four-dimensional phase-space $(\bm
R, \dot{\bm R})$ reduces to
\begin{eqnarray}
  \dot X& = & - X - Z^{-1}\, \left(h X^2 -Y^2\right) + \eta_1(s)\,,
  \label{eqn:eomx} \\ \dot Y & = & - Y - (1+h)\, Z^{-1}\, X\,Y
  +\eta_2(s)\,,
  \label{eqn:eomy} \\ \dot Z & = & (1-h)\, X\,. \label{eqn:eomz}
\end{eqnarray}
Now the dots denote derivatives with respect to $s = t/\tau$; $\eta_1$
and $\eta_2$ are independent white noises with variances $2\,\St(L)$
and $2\,(1+2h)\,\St(L)$, respectively; $\St(L) = D_1\,\tau /
L^{2(1-h)}$ is the Stokes number associated with the system
size. Periodic boundary conditions in physical space amount to
considering reflective boundary conditions at $Z=1$; $Y$ is ensured to
remain positive by reflective boundary conditions at $Y=0$. Rescaling
$|\bm R|$ with $\lambda$, and thus $Z$ with $\lambda^{1-h}$ leads to
transform $X$ and $Y$ to $\lambda^{1-h}X$ and $\lambda^{1-h}Y$ in
order to confine the scaling factor in the noise.  This again amounts
to considering the same dynamics with a scale-dependent Stokes number
$S(\lambda L)$. The system (\ref{eqn:eomx})-(\ref{eqn:eomz}) can be
efficiently implemented numerically and was used to produce the data
described in this Letter.

\begin{figure}[t!]
  \centerline{\includegraphics[width=0.48\textwidth]{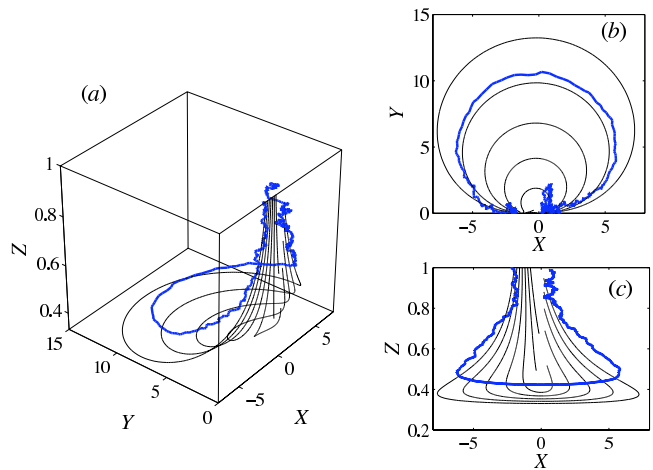}}
  \vspace{-10pt} \caption{Phase-space picture of the system
  (\ref{eqn:eomx})-(\ref{eqn:eomz}) for $h=7/10$. The thin smooth
  lines represent effects of the drift.  A random trajectory of the
  system with $\St(L) = 1$ is shown as a bold line; it performs a
  large loop from $X<0$ to $X>0$. (a) The full $(X,Y,Z)$ space, (b)
  and (c) projections in the $Z=0$ and $Y=0$ planes,
  respectively. \vspace{-10pt}}
  \label{fig:randomtraj}
\end{figure}

Figure \ref{fig:randomtraj} sketches the dynamics in the $(X,Y,Z)$
space. The line $X\!=\!Y\!=\!0$ (its physical meaning is that the
particles have relaxed to the same velocity staying at an arbitrary
distance) acts as a stable fixed line for the drift terms in
(\ref{eqn:eomx})-(\ref{eqn:eomz}). A typical trajectory spends a long
time diffusing around this line, until the noise realization becomes
strong enough to escape from its neighborhood. When this happens with
$X\!>\!0$, the quadratic terms in the drift drive the trajectory back
to the stable line. On the contrary, if $X\!<\!0$ and $hX^2\!+\! XZ\!
-\!Y^2\! <\! 0$, the drift accelerates the trajectory towards larger
negative values of $X$. Then $Z$ decreases until the quadratic terms
in (\ref{eqn:eomx})-(\ref{eqn:eomy}) become dominant. The trajectory
then loops back in the $(X,Y)$-plane, approaching the stable line from
its right. These loops are the events during which $Z$ (and hence the
inter-particle distance $R$) becomes substantially small.

As we now show, these loops are responsible for power-law tails in the
probability density function (PDF) of the dimensionless velocity
differences $X$ and $Y$.  This behavior can be understood by extending
to the rough case the arguments developed in \cite{bch06} for smooth
flows.  For this we consider the cumulative probability $P^<(x) = {\rm
Pr}\,(X < x)$ with $x\!\ll\!-1$, which can be estimated as the product
of (\textit{i}) the probability to start a sufficiently large loop
that reaches values more negative than $x$ and (\textit{ii}) the
fraction of time spent by the trajectory at $X < x$. To estimate these
two contributions, we assume that within a distance of order unity
from the line $X=Y=0$ the quadratic terms in the drift are negligible,
so that $X$ and $Y$ behave as two independent Ornstein--Uhlenbeck
processes.  At larger distances we retain only the quadratic terms
responsible for the loops.

Within this simplified dynamics, a loop is initiated at a time $s_0$
for which $X_0\!=\!X(s_0)\!<\! -1$ and $Y_0\!=\!Y(s_0)\!\ll
\!|X_0|$. The maximum distance from the stable line is attained at a
time $s^\ast$ for which $X(s^\ast)$ is of the order of $-Y(s^\ast)$,
that is when $X(s^\ast)\! =\!  -\beta\,Y(s^\ast)$, $\beta$ being an
arbitrary constant. When the noise is neglected, one straightforwardly
obtains: ${Y(s)}/{X(s)} = {Y_0\,Z_0}/{\{X_0\,Z_0+(1\!-\!{\rm
e}^{s_0-s}) \, [X_0^2+Y_0^2]\}}$ and the radius of the loop can be
estimated as
\begin{equation}
  |X(s^\ast)| = \frac{\beta\,[X_0+Z_0]\,X_0^{h}}
  {[1+\beta^2]^{(1+h)/2}}\, Y_0^{-h}\,.
\end{equation}
For the trajectory to reach values $X<x\ll -1$, the radius has to be
larger than $-x$, and thus $Y(s_0)$ has to be smaller than
$|x|^{-1/h}$. The joint PDF of $X$ and $Y$ at time $s_0$ is given by
the dynamics close to the line $X\!=\!Y\!=\!0$.  As it is finite for
$Y\!\to\! 0$, the contribution (\textit{i}) can be estimated to be
$\propto |x|^{-1/h}$. The contribution (\textit{ii}) can be obtained
as follows. Far from the stable line, the dynamics can be approximated
by the deterministic part, hence the fraction of time spent at $X<x$
is $\propto Y_0 \propto |x|^{-1/h}$. Putting together the two
contributions, we obtain that $P^<(x)\propto |x|^{-2/h}$ when
$x\ll-1$. The negative tail of the PDF of the longitudinal velocity
difference $X$ behaves as a power-law $\propto |x|^{-\alpha}$ with
$\alpha = 1+2/h$.
This gives $\alpha=3$ for smooth flows ($h=1$) as previously derived
\cite{bch06}.  During the large loops, the trajectories equally reach
large {\em positive} values of $X$ and of $Y$. This gives again a
fraction of time $x^{-1/h}$ spent at both $X$ and $Y$ larger than $x
\gg 1$. Hence, the PDF of both the longitudinal and the transversal
velocity differences have algebraic left and right tails.

\begin{figure}[t!]
  \centerline{\includegraphics[height=0.35\textwidth]{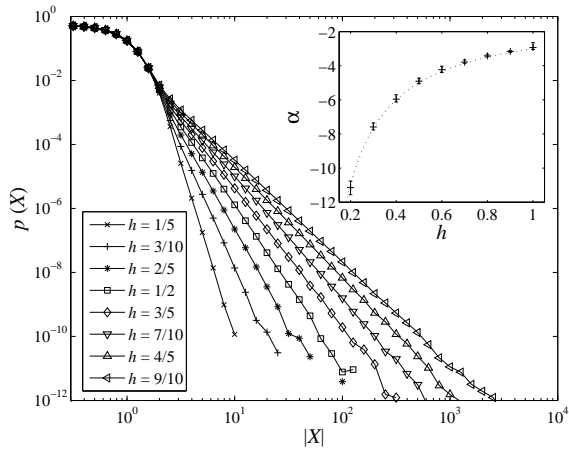}}
  \vspace{-10pt}
  \caption{PDF of $X$ in log-log coordinates for $\St(L) = 1$ and
    various values of $h$. In all cases, power-law tails are
    observed. Inset: exponent $\alpha$ of the algebraic tail as a
    function of the fluid velocity H\"older exponent $h$; the
    theoretical prediction is represented as a dotted
    line. \vspace{-10pt} }
  \label{fig:tailpdfX-diffh}
\end{figure}
As shown in Fig.~\ref{fig:tailpdfX-diffh}, the presence of power-law
tails in the PDF is confirmed numerically, with perfect agreement
between the measured values of $\alpha$ and the prediction $\alpha =
1+2/h$ (see inset).  Let us comment on the $h$ dependence of this
exponent $\alpha$.  The large loops responsible for the algebraic
tails correspond to events in which particles approach each other very
closely; they are the basic mechanism of particle clustering.  The
probability to enter such loops decreases significantly when $h\to 0$.
Moreover, it is straightforward to check from
(\ref{eqn:eomx})-(\ref{eqn:eomz}) that during the loops $Z(s) \propto
Z_0^h$ when $Z_0\ll 1$. Hence it gets less and less probable to reach
smaller values of $Z$ as $h$ decreases. Combined together, these two
effects explain why particle clustering is weakened in rough velocity
fields and why it is more efficient in smooth flows.

The change of variables (\ref{eq:changevar}) can be equally applied in
three dimensions, leading to a dynamics different from
(\ref{eqn:eomx})-(\ref{eqn:eomz}). Therefore understanding to what
extent the above findings extend to the 3D case remains an open
question; work in this direction is under development.

To conclude, let us comment on the implications of this work on the
study of heavy particles in realistic turbulent flows.  There,
particle clustering is simultaneously due to ejection from eddies and
to a dissipative dynamics. The considered model flow isolates the
latter effect. It is probable that power-law tails for velocity
differences can be present in realistic settings as well.  However, it
is not clear if the results on clustering are affected by the presence
of persistent structures: particle ejection from eddies may form voids
and thus very strong inhomogeneities in the particle
distribution~\cite{bdg04}.  This could overtake dissipative-dynamics
mechanisms.\\
\indent We acknowledge useful discussions with L.\ Biferale,
G.~Falkovich, K.~Gaw\c{e}dzki, A.~Lanotte, S.\ Musacchio, and F.\
Toschi.  This work has been partially supported by the EU network
HPRN-CT-2002-00300 and by the French-Italian Galileo program
``Transport and dispersion of impurities in turbulent flows''. The
stay of R.H.\ in Nice was supported by the Zeiss-Stiftung and the
DAAD.

\bibliography{roughletter}

\end{document}